\documentclass[english,prl,showpacs,twocolumn]{revtex4}
\usepackage[T1]{fontenc}
\usepackage[latin1]{inputenc}
\usepackage{babel}
\usepackage{graphics}

\makeatletter

\providecommand{\LyX}{L\kern-.1667em\lower.25em\hbox{Y}\kern-.125emX\@}

\usepackage[T1]{fontenc}
\usepackage[latin1]{inputenc}
\usepackage{graphics}

\makeatletter

\usepackage{revsymb}

\makeatother

\makeatother
\begin{document}

\title{Universal Attractors of Reversible Aggregate-Reorganization Processes }

\author{Stefan Großkinsky\( ^{1} \), Marc Timme\( ^{2} \), and Björn Naundorf\( ^{2} \)}

\affiliation{\( ^{1} \)Zentrum Mathematik, TU München, 80290 München, Germany
\\
 \( ^{2} \)Max-Planck-Institut für Strömungsforschung, 37073 Göttingen,
Germany}

\begin{abstract}
We analyze a general class of reversible aggregate-reorganization
processes. These processes are shown to exhibit globally attracting
equilibrium distributions, which are \textit{universal}, i.e.\ identical
for large classes of models. Furthermore, the analysis implies that
for studies of equilibrium properties of \textit{any} such process,
computationally expensive reorganization dynamics such as random walks
can be replaced by more efficient, yet simpler methods. As a particular
application, our results explain the recent observation of the formation
of similar fractal aggregates from different initial structures by
diffusive reorganization {[}Filoche and Sapoval, \textit{Phys.\ Rev.\ Lett.}\textbf{\ 85},
5118 (2000){]}. 
\end{abstract}

\pacs{61.43.Hv, 64.60.Cn, 89.75.Fb}

\maketitle
Many complex systems like microbial colonies, colloidal aggregates,
or particle-adsorbate structures are composed of a large number of
similar units \cite{Bunde_et_al}. For the growths of such particle
clusters, the model of diffusion-limited aggregation (DLA) has been
introduced \cite{Witten} and intensively investigated in several
generalizations \cite{Bunde_et_al, Meakin88, Erzan}. In these non-equilibrium
models, particles are permanently injected into the system and stick
\textit{irreversibly} to an existing cluster which grows towards a
fractal structure. In real systems, however, particles are often rearranged,
dynamically transforming the aggregate structure. This rearrangement
is appropriately modelled by processes exhibiting \textit{reversible}
dynamics \cite{Bunde_et_al,Meakin88,FilochePRL}.

Recently, Filoche and Sapoval have proposed a specific model of diffusion-reorganized
aggregation (DRA) \cite{FilochePRL} in which particles detach from
the boundary of a given cluster of fixed size and reaggregate after
a random walk on the underlying square lattice. In numerical simulations,
they observe the formation of fractal structures similar to those
found in various irreversible models. The DRA process transforms different
initial aggregates, e.g.\ simple one-dimensional and two-dimensional
structures, into statistically similar fractal aggregates after many
reorganization steps.

In this Letter, we analyze a general class of reversible processes
that dynamically reorganize a fixed number of particles constituting
one connected aggregate on some lattice. Particles are allowed to
reorganize by any reversible process that retains a connected aggregate,
including diffusion as a particular case. We prove that all these
processes exhibit equilibrium distributions that are globally attracting.
Our general mathematical perspective reveals that such attractors
are \textit{universal}, i.e.\ independent of the dynamical model details:
In discrete-time models, this distribution only depends on the mechanism
of particle disaggregation but is independent of the subsequent particle
repositioning dynamics. In particular, this implies that ordinary
diffusion and different mechanisms like surface diffusion \cite{Bunde_et_al}
or Lévy-flights \cite{Levy} are equivalent in this context. Furthermore,
we show that in continuous-time models equilibria are independent
of both the mechanism of repositioning and that of disaggregation.
On a \( d \)-dimensional cubic lattice, if all connected aggregates
are accessible by the repositioning dynamics, this leads to the uniform
distribution that is known as the ensemble of lattice animals (cf.\ \cite{Janse, BotetAndWessel, Lubensky_et_al}),
currently also investigated in number theory under the name polyominoes
(\cite{Golomb_et_al} and refs. therein).

In addition to these analytical results, we present strong numerical
evidence that, for sufficiently large aggregates, there is no significant
difference between equilibrium distributions obtained from discrete- or continuous-time
models. Our results are valid for a general class of processes, including
reorganization of aggregates consisting of different kinds of particles
as well as repositioning of whole clusters instead of single particles.
Here, for simplicity of presentation, we focus on processes in which
single, identical particles are reorganized.

\begin{figure}
{\centering \resizebox*{8cm}{!}{\includegraphics{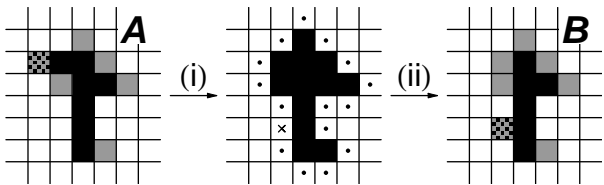}} \par}

\caption{{\small Model of a specific reversible aggregate-reorganization process:
One particle (checkered) is selected from a set of free particles
\protect\protect\( F(A)\protect \protect \) (gray) of aggregate} \textsf{\small  \protect\protect\( A\protect \protect \)}
{\small according to rule (i) and detaches from the aggregate. By
a reversible process, e.g.\ diffusion, it is repositioned at one (\protect\protect\( \times \protect \protect \))
of the free aggregation sites \protect\protect\( D(A,B)\protect \protect \)
(\protect\protect\( \bullet \protect \protect \)) of the intermediate
object according to rule (ii). Reversibility requires that the deposited
particle (checkered) is again a free particle in \protect\protect\( F(B)\protect \protect \)
(gray) and may be selected in the next reorganization step.\label{fig:rules}}}
\end{figure}

We consider reorganization processes of aggregates -- certain connected
collections of a fixed number \( N \) of particles on some lattice.
Starting from an aggregate \( A \), one step of the discrete-time
dynamics is defined by the following two rules (Fig.\ \ref{fig:rules}):

\begin{itemize}
\item[(i)] Randomly select a particle from a set of free particles \( F(A) \)
with equal probability, where \( F(A) \) is chosen such that the
aggregate remains connected. 
\item[(ii)] Reposition this particle by a reversible process at some site
\( x \) of possible aggregation sites \( D(A,B) \) of an intermediate
object. By reversibility, this particle is again a free particle of
the new aggregate \( B \). 
\end{itemize}
The above two rules are repeatedly applied to the evolving aggregate.
Note that physically the step-by-step application of the two rules
reflects a separation of time scales that is realized in many processes
\cite{Bunde_et_al}: The disaggregation of a particle occurs much
slower than its repositioning. To model a specific reorganization
process, the sets \( F(A) \) of free particles (i), the repositioning
(ii) as well as the underlying lattice can be defined as desired as
long as every aggregate under consideration can be reached iterating
the above two rules.

To be specific we focus on \( d \)-dimensional cubic lattices, but
the analysis can be analogously applied for arbitrary lattices. We
fix aggregates on a lattice of \( (N+1)^{d} \) sites with periodic
boundary conditions and view translated aggregates of the same shape
as different states. A given process is identified as a Markov chain
\cite{Norris} with state space \( S_{N} \) consisting of all connected
aggregates of \( N \) particles, which have at least one free particle,
i.e.\ non-empty \( F(A) \). The state of this discrete-time process
is then described by the distribution \( p_{t}(A) \), giving the
probability that after \( t \) reorganization steps the aggregate
\( A \) is present. Starting with initial distribution \( p_{0}(A) \),
a transition matrix \( T \) specifies the time evolution

\begin{equation}
\label{eq:matrix_{e}quation}
\mathbf{p}_{t}=T\mathbf{p}_{t-1}=T^{t}\mathbf{p}_{0}
\end{equation}
 of the state vector \( \mathbf{p}_{t}:=(p_{t}(A_{1}),\ldots ,p_{t}(A_{|S_{N}|}))^{\mathrm{T}} \).
As \( S_{N} \) is finite, the long-time (\( t\rightarrow \infty  \))
behaviour of the discrete-time Markov chain is determined by a steady
state distribution \( \mathbf{p}_{d}^{*}=T\mathbf{p}_{d}^{*} \).
If the chain is irreducible, i.e.\ every aggregate \( A\in S_{N} \)
can be reached by the process from every other, \( \mathbf{p}_{d}^{*} \)
is the global attractor towards which every initial distribution converges
\cite{Norris}. This uniqueness property depends on the choice of
free particles \( F(A) \). In general, one reasonably restricts the
state space \( S_{N} \) to an appropriate subset of aggregates consistent
with \( F(A) \). For instance, if \( F(A) \) is the set of all one-bond
particles \cite{BotetAndWessel}, only tree-like aggregates are considered.
In such cases our analysis applies to the respective irreducible subset
of \( S_{N} \) (also simply denoted by \( S_{N} \) in the following).
With this convention, the stationary distribution \( \mathbf{p}_{d}^{*} \)
is the unique attractor for every initial distribution \( \mathbf{p}_{0} \)
on \( S_{N} \). We now specify the transition matrix \( T \) which
completely determines the dynamics of the process.

If a one-step transition from aggregate \( A \) to \( B \) is possible,
denoted by \( A\leftrightarrow B \), the particle which is disaggregated
and its reaggregation site are determined uniquely \cite{aa}. Thus
the underlying lattice does not explicitly enter the analysis. The
transition probabilities are given by the matrix elements \begin{equation}
\label{eq:matrix_{e}lement}
T_{BA}=\left\{ \begin{array}{ll}
\frac{1}{|F(A)|}\, r_{BA} & \, \, \, \mbox {if}\, A\leftrightarrow B\\
0 & \, \, \, \mbox {otherwise}
\end{array}\right. 
\end{equation}
 which, if non-zero, consist of two factors: According to rule (i)
a free particle is randomly selected from \( F(A) \) with uniform
probability \( 1/|F(A)| \). Then, the particle is repositioned at
the intermediate aggregate with probability \( r_{BA} \) to obtain
aggregate \( B \) according to rule (ii). The reversibility of the
repositioning procedure is characterized by the symmetry \( r_{BA}=r_{AB} \).
Thus, the dynamics is described by the master equation\begin{eqnarray}
p_{t+1}(A)-p_{t}(A) & = & \sum _{B\in S_{N}}\left[ T_{AB}\, p_{t}(B)-T_{BA}\, p_{t}(A)\right] \nonumber \\
 & = & \sum _{B\in S_{N}}\left[ \frac{p_{t}(B)}{|F(B)|}-\frac{p_{t}(A)}{|F(A)|}\right] r_{AB}\, .\label{eq:master_discr} 
\end{eqnarray}
 A steady state solution of this equation is obviously given by the
equilibrium distribution \begin{equation}
\label{eq:discrete_equilibrium}
p_{d}^{*}(A)=|F(A)|/\sum _{B\in S_{N}}|F(B)|,
\end{equation}
 which fulfills the condition of detailed balance, i.e.\ \( \frac{p_{d}^{*}(B)}{|F(B)|}=\frac{p_{d}^{*}(A)}{|F(A)|} \)
for all \( A,B\in S_{N}. \) From the above we know that \( \mathbf{p}_{d}^{*} \)
is the unique attractor of the dynamics (\ref{eq:master_discr}).
Hence, in equilibrium the probability to observe an aggregate \( A \)
is proportional to its number \( |F(A)| \) of free particles.

It seems natural to realize repositioning by a random walk of the
disaggregated particle, because it models diffusion in real systems
\cite{FilochePRL, BotetAndWessel}. This random walk determines the
transition probabilities \( r_{AB}=r_{BA} \) between aggregates \( A \)
and \( B \). However, independent of the actual values of \( r_{AB} \),
and hence the repositioning procedure (ii), the process leads to the
\textit{same equilibrium distribution} \( \mathbf{p}_{d}^{*} \).
Thus, to study the equilibrium properties of the reorganization process,
one can replace the random walk in numerical simulations: For instance,
after selection of a particle, it may just be repositioned at every
aggregation site with equal probability \( r_{AB}=1/|D(A,B)| \),
where \( D(A,B) \) is the set of possible aggregation sites of the
intermediate aggregate (cf.\ Fig.\ \ref{fig:rules} and ref.\ \cite{Janse}).
This straightforward procedure is much simpler to implement than random
walks and significantly reduces the computational effort. Furthermore,
one expects a faster convergence to the equilibrium distribution \cite{Guruswami}:
While for diffusive repositioning the particle re-aggregates with
high probability near or even at its disaggregation site, the alternative
process leads to a fast spreading of particles so that the state space
is likely to be sampled faster. Moreover, diffusing particles in \( d\geq 2 \)
dimensions have an infinite average return time leading to computational
complications \cite{FilochePRL, BotetAndWessel} that can be avoided
using alternative processes.

Up to now, we have considered discrete-time evolution rules, which
arise when performing computer simulations. However, for real reorganization
processes, continuous-time modelling, where the free particles \( F(A) \)
disaggregate independently at a rate \( \gamma  \), often seems more
appropriate. This leads to a continuous-time Markov chain with completely
symmetric transition rates \( \gamma \, r_{AB} \) occurring in the
master equation \begin{equation}
\label{eq:master_cont}
\frac{d}{dt}p_{t}(A)=\sum _{B\in S_{N}}\, \gamma \, \left[ p_{t}(B)-p_{t}(A)\right] r_{AB}\, .
\end{equation}
 Thus, the attractor of these continuous-time processes is \begin{equation}
\label{eq:continuous_equilibrium}
p_{c}^{*}(A)=1/|S_{N}|.
\end{equation}
 This is a universal equilibrium distribution which is \textit{independent
of all dynamical model details}. It is important to note that, although
\( F(A) \) does not enter (\ref{eq:continuous_equilibrium}), it
implicitly determines the set \( S_{N} \) of accessible aggregates.
In the case of \( d \)-dimensional cubic lattices and for every choice
of \( F(A) \) for which the process can reach all possible connected
aggregates of \( N \) particles (cf.\ our simulations below), this
uniform distribution \( \mathbf{p}_{c}^{*} \) is known as the ensemble
of lattice animals \cite{Lubensky_et_al, Golomb_et_al}.

\begin{figure}
{\centering \resizebox*{7cm}{!}{\includegraphics{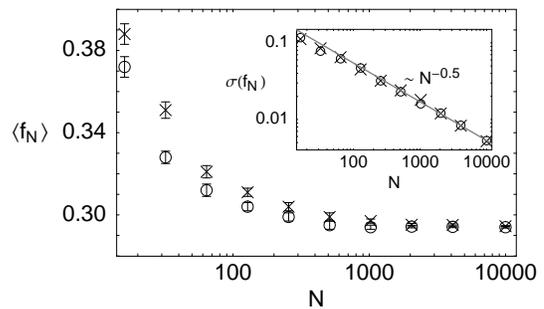}} \par}

\caption{{\small Comparison of average fractions \protect\protect\( \left\langle f_{N}\right\rangle \protect \protect \)
of free particles: discrete-time (\protect\protect\( \times \protect \protect \))
and its associated continuous-time model (\protect\protect\( \bigcirc \protect \protect \)).
Error bars denote standard error of sample mean. Inset: Widths \protect\protect\( \sigma (f_{N})\protect \protect \)
for both time scales. \label{fig:discrete_vs_continuous}}}
\end{figure}

The question arises, how relevant the differences between discrete-
and continuous-time modelling are. To answer this question, we consider
the fraction \begin{equation}
\label{eq:fraction}
f_{N}(A)=|F(A)|/N
\end{equation}
 of free particles of an aggregate \( A \) of \( N \) particles.
This observable determines the difference in the equilibrium distributions
of discrete-time and their associated continuous-time models because
it directly enters (\ref{eq:discrete_equilibrium}) in the form \( p^{*}_{d}(A)=f_{N}(A)/\sum _{B}f_{N}(B) \).
In equilibrium, the average fractions of free particles obviously
satisfy the inequality \begin{eqnarray}
\left\langle f_{N}\right\rangle _{\mathbf{p}^{*}_{d}}=\sum _{A}f_{N}(A)p^{*}_{d}(A)=\frac{\sum _{A}f^{2}_{N}(A)}{\sum _{B}f_{N}(B)} & \geq \left\langle f_{N}\right\rangle _{\mathbf{p}^{*}_{c}}\qquad \label{eq:Jensen} 
\end{eqnarray}
 because \( \left\langle f_{N}\right\rangle _{\mathbf{p}^{*}_{c}}=\sum _{B}f_{N}(B)/|S_{N}| \)
by definition. Nevertheless, Monte-Carlo simulations of a specific
process \cite{simulations} show that the averages \( \left\langle f_{N}\right\rangle  \)
clearly converge towards the same value for discrete- and continuous-time
models (Fig.\ \ref{fig:discrete_vs_continuous}). 
Differences become insignificant already for moderately sized aggregates. 

This is quantified by considering the widths of the equilibrium distributions
\( \sigma (f_{N}):=\left\langle \left( f_{N}-\left\langle f_{N}\right\rangle \right) ^{2}\right\rangle ^{\frac{1}{2}} \)
(Fig.\ \ref{fig:discrete_vs_continuous}, inset). We find a power
law decay \( \sigma (f_{N})\sim N^{-\alpha } \) with \( \alpha =0.50\pm 0.02 \)
for both time scales as would be expected for large random structures.
This has important consequences: Calculations similar to the above
(\ref{eq:Jensen}) show immediately that \begin{equation}
\label{eq:arbitrary_scaling}
\left\langle g_{N}\right\rangle _{\mathbf{p}^{*}_{d}}=\left\langle g_{N}\right\rangle _{\mathbf{p}^{*}_{c}}\left( 1+\mathcal{O}(N^{-\alpha })\right) 
\end{equation}
for \textit{every} observable \( g_{N}(A) \). Thus, for sufficiently
large aggregates, measurements of \textit{arbitrary} observables,
and, equivalently, the equilibrium distributions, in discrete- and
continuous-time models cannot be distinguished. This reflects the
fact that the number \( |F(A)| \) of free particles of almost all
aggregates \( A\in S_{N} \) is close to its uniform ensemble average
\( \left\langle |F(A)|\right\rangle  \) for sufficiently large \( N \).
Since this fact is of purely configurational origin, we expect a similar
convergence (\ref{eq:arbitrary_scaling}) for other models, also in
\( d>2 \) dimensions.

\begin{figure}
{\centering \resizebox*{6cm}{!}{\includegraphics{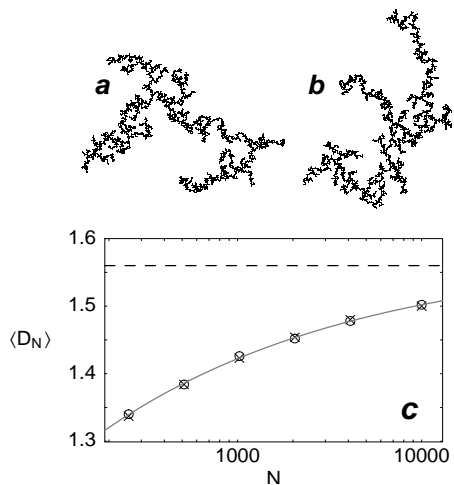}} \par}

\caption{{\small (a,b) Equilibrium fractal aggregates of \protect\protect\( N=10000\protect \protect \)
particles, started from a \protect\protect\( 100\times 100\protect \protect \)
square, after \protect\protect\( 10^{9}\protect \protect \) particle
rearrangements \cite{simulations}. (a) Discrete-time model, (b) associated
continuous-time model. (c) Box-counting dimensions \protect\protect\( \left\langle D_{N}\right\rangle \protect \protect \)
for both time-scales are indistinguishable and approach from below
\cite{scaling} the fractal dimension \protect\protect\( D\approx 1.56\protect \protect \)
(dashed line) \cite{Lubensky_et_al, Golomb_et_al}. \label{fig:two_fractals}} }
\end{figure}

As one important observable, we have studied the fractal dimensions
of aggregates from both processes. Sample aggregates as well as box-counting
dimensions \( \left\langle D_{N}\right\rangle  \) are shown in Fig.\ \ref{fig:two_fractals}.
Even for moderately sized aggregates, no significant differences have
been observed. With increasing aggregate sizes, the \( \left\langle D_{N}\right\rangle  \)
approach from below \cite{scaling} the well-estimated fractal dimension
\( D\approx 1.56 \) \cite{Lubensky_et_al, Golomb_et_al} of square-lattice
animals.

The above results predict that the uniform distribution among all
connected aggregates which can be reached by a specific process is
the globally attracting, unique equilibrium distribution for a wide
class of models. In particular, our results explain the recent observation
that fractal aggregates are reached from arbitrary initial structures
by a certain diffusive reorganization process \cite{FilochePRL}.
Nevertheless, our results strongly indicate that the equilibrium aggregates
from that study are in the same universality class as lattice animals
and thus have a fractal dimension \( D\approx 1.56 \) and not, as
suggested, \( D=1.74\pm 0.02 \) \cite{FilochePRL}. A possible explanation
of this discrepancy is that simulations in that study did not reach
equilibrium at the time of measurement, despite a huge number of simulation
steps. In fact, in our simulations \cite{simulations} aggregates
of \( N=10000 \) particles only reached equilibrium after about
\( 2\times 10^{8} \) particle rearrangements. 

It is important to note, that in certain systems such a slow convergence may 
imply that the equilibrium state is not reached during the time of real 
or numerical experiments (cf.\ \cite{FilochePRL}). Thus although an 
equilibrium distribution is guaranteed to be reached, the long-term behavior 
of such systems may be governed by transient states with long lifetimes and 
non-universal statistical properties.

In summary, we have shown that large classes of reversible aggregate-reorganization
processes possess universal attractors which are independent of the
model details but may in general depend on whether discrete- or continuous-time
modelling is used. Furthermore, our numerical investigations indicate
that statistical properties are essentially identical for both time
scales. These results imply that commonly used diffusion modelling
can be replaced by computationally fast methods which in addition
lead to a faster convergence towards the (identical) equilibrium distribution.
In general, one specific process of aggregate-reorganization can be
investigated in a number of different alternative models. Which alternative
is used can be freely chosen depending on which one is more convenient
to study analytically or by computer simulations.

In particular, our results explain the recently discovered {}``statistical
attractors'' in diffusive reorganization processes \cite{FilochePRL}
but predict a different final equilibrium state. More generally, the
results can serve as a basis for investigating equilibrium statistical
properties of a general class of aggregate-reorganization processes.
Reversible reorganization is always characterized by the symmetry
\( r_{AB}=r_{BA} \), which, in turn, has been the main requirement
for universality. In particular, aggregates consisting of different
kinds \( \alpha  \) of particles disaggregating with particle-specific
rates \( \gamma _{\alpha } \) are also described because a transition
\( A\leftrightarrow B \) determines the disaggregating particle uniquely.
Moreover, the theory may be generalized to reorganization of whole
sub-clusters rather than just single particles (cf.\ also \cite{Meakin88, Janse}).
Hence more complex models of aggregate-reorganization can also be
captured.

We have imposed only a small number of requirements on the dynamics
which appear to be natural assumptions in many physical situations,
in particular reversibility of repositioning and separation of time-scales.
Thus our main results might not only be of importance to aggregate-reorganization
processes. We hope that the basic ideas can also find applications
in studies of other complex systems exhibiting reversible reorganization.

We thank M. Biehl, D. Brockmann, T. Geisel, L. Hufnagel, T. Kottos,
and M. Prähofer for useful discussions. This work has been supported
by the Max Planck Society.


\begin{thebibliography}{10}
\bibitem{Bunde_et_al}{\small A. Bunde and S. Havlin (eds.),} \textit{\small Fractals and
Disordered Systems}{\small , (Springer, Berlin, 1991); T. Vicsek,}
\textit{\small Fractal Growth Phenomena} {\small (World Scientific,
Singapore, 1992); P. Meakin,} \textit{\small Fractals,} \textit{\small scaling
and growth far from equilibrium} {\small (Cambridge Univ. Press, Cambridge,
1998). \par{}}{\small \par}
\bibitem{Witten}{\small T.A. Witten and L.M. Sander, Phys. Rev. Lett.} \textbf{\small 47}{\small ,
1400 (1981).\par{}}{\small \par}
\bibitem{Meakin88}{\small P. Meakin, in} \textit{\small Phase Transitions and Critical
Phenomena}{\small , Vol. 12, edited by C. Domb and J.L. Lebowitz (Academic Press,
London, 1988).\par{}}{\small \par}
\bibitem{Erzan}{\small A. Erzan, L. Pietronero, and A. Vespignani, Rev. Mod. Phys.}
\textbf{\small 67}{\small , 545 (1995).\par{}}{\small \par}
\bibitem{FilochePRL}{\small M. Filoche and B. Sapoval, Phys. Rev. Lett.} \textbf{\small 85}{\small ,
5118 (2000).\par{}}{\small \par}
\bibitem{Levy}{\small P. Meakin, Phys. Rev. B} \textbf{\small 29}{\small , 3722
(1984).\par{}}{\small \par}
\bibitem{Janse}{\small E.J. Janse van Rensburg and N. Madras, J. Phys. A} \textbf{\small 25}{\small ,
303 (1992).\par{}}{\small \par}
\bibitem{BotetAndWessel}{\small R. Botet and R. Jullien, Phys. Rev. Lett.} \textbf{\small 55}{\small ,
1943 (1985); R. Wessel and R.C. Ball, Phys. Rev. A} \textbf{\small 45}{\small ,
R2177 (1992).\par{}}{\small \par}
\bibitem{Lubensky_et_al}{\small T.C. Lubensky and J. Isaacson, Phys. Rev. A} \textbf{\small 20}{\small ,
2130 (1979); A. L. Stella} \textit{\small et al.}{\small , Phys. Rev.
Lett.} \textbf{\small 69}{\small , 3650 (1992); D. Stauffer and A.
Aharony,} \textit{\small Introduction to Percolation Theory} {\small (Taylor
and Francis, London, 1994). \par{}}{\small \par}
\bibitem{Golomb_et_al}{\small S. Golomb,} \textit{\small Polyominoes: Puzzles, Patterns,
Problems, and Packings}{\small , 2nd ed. (Princeton Univ. Press, Princeton,
1994); I. Jensen and A.J. Guttmann, J. Phys. A} \textbf{\small 33}{\small ,
L257 (2000). \par{}}{\small \par}
\bibitem{Norris}{\small J.R. Norris,} \textit{\small Markov chains}{\small , (Cambridge
Univ. Press, Cambridge, 1997).\par{}}{\small \par}
\bibitem{aa}{\small Uniqueness is guaranteed only if \( B\neq A \). If \( B=A \),
any particle from \( F(A) \) can disaggregate and must then reaggregate
at the same site. This is easily included into the definition of \( r_{AA} \).
\par{}}{\small \par}
\bibitem{Guruswami}{\small V. Guruswami,} \textit{\small Rapidly Mixing Markov Chains,}
{\small Survey (MIT Laboratory for Computer Science, Cambridge, 2000).
\par{}}{\small \par}
\bibitem{simulations}{\small Monte-Carlo simulations: square lattice, \( F(A) \) =\{all
particles which, if removed, keep the aggregate connected\}. Expectation
values estimated from at least \( 250 \) equilibrium realizations
of aggregates (depending on \( N \)). \par{}}{\small \par}
\bibitem{scaling}{\small Finite size scaling \( D_{\infty }-\left\langle D_{N}\right\rangle \sim N^{-\nu } \)
with \( \nu =0.33\pm 0.04 \) leads to a dimension estimate of \( D_{\infty }=1.56\pm 0.02 \).\par{}}\end{thebibliography}
\end{document}